\documentclass[conference,usenames,dvipsnames,x11names,table,10pt,compsocconf]{IEEEtran}
\IEEEoverridecommandlockouts
\usepackage[abbreviations]{foreign}
\usepackage[binary-units]{siunitx}
\usepackage[utf8]{inputenc}
\usepackage{graphicx}
\usepackage{xcolor}  
\usepackage{amsmath,amssymb}
\usepackage{algorithm,}
\usepackage[noend]{algpseudocode}
\usepackage{ifthen}
\usepackage{multicol}
\usepackage{enumitem}
\usepackage{color}
\usepackage{listings}
\usepackage{pifont}
\usepackage{multirow}
\usepackage[hyphens,spaces,obeyspaces]{url}
\usepackage{hyperref}
\usepackage{cleveref} 
\usepackage{amsthm}
\usepackage{array}
\usepackage{float}
\usepackage{booktabs}
\usepackage{subcaption}
\usepackage[backend=bibtex,style=numeric,natbib=true]{biblatex}
\newboolean{showcomments}
\setboolean{showcomments}{true}
\ifthenelse{\boolean{showcomments}}
{ \newcommand{\mynote}[3]{
   \fbox{\bfseries\sffamily\scriptsize#1}
   {\small$\blacktriangleright$\textsf{\emph{\color{#3}{#2}}}$\blacktriangleleft$}}}
{ \newcommand{\mynote}[3]{}}

\newcommand{\system}{\textsc{Abeona}\xspace}
\newcommand{\sys}{\system}

\bibliography{compact}

\renewbibmacro{finentry}{%
    \iffieldequalstr{entrykey}{arm_trustzone}
    {\finentry\newpage}
    {\finentry}}

\begin{document}
\title{\sys: an Architecture for Energy-Aware\\Task Migrations from the Edge to the Cloud}
\date{}

\author{
\IEEEauthorblockN{
Isabelly Rocha,\IEEEauthorrefmark{1}
Gabriel Vinha,\IEEEauthorrefmark{2}
Andrey Brito,\IEEEauthorrefmark{2}
Pascal Felber,\IEEEauthorrefmark{1}
Marcelo Pasin\IEEEauthorrefmark{1} and
Valerio Schiavoni\IEEEauthorrefmark{1}
}
\IEEEauthorblockA
{\IEEEauthorrefmark{1}University of Neuchâtel, Switzerland} 
{\IEEEauthorrefmark{2}Universidade Federal de Campina Grande, PB, Brazil}
}

\maketitle
\thispagestyle{plain}
\pagestyle{plain}

\begin{abstract}
This paper presents our preliminary results with \sys, an edge-to-cloud architecture that allows migrating tasks from low-energy, resource-constrained devices on the edge up to the cloud.
Our preliminary results on artificial and real-world datasets show that it is possible to execute workloads in a more efficient manner energy-wise by scaling horizontally at the edge, without negatively affecting the execution runtime.
\end{abstract}
\vspace{-10pt}
\section{Introduction}
\label{sec:introduction}
Despite being in its infancy, the Internet of Things (IoT) is nowadays gathering a lot of momentum in industry and academia.
By the year 2025, it is prospected \cite{safaei2017reliability} that each individual will have more than 9 smart devices.
These devices cooperate and collaborate, perhaps in untrusted environments, at the \emph{edge} of the network.
Applications running at the edge can for instance react quickly to local data, with clear latency advantages when compared to cloud-based applications.
Edge devices can also conveniently aggregate, filter or anonymise local data before making it available to the cloud, adding more privacy to the applications.
In this context, we study how scheduling, placement and migration criterias for task-based workloads are affected.

A common practice is to deploy groups of small devices in connected clusters directly reachable by closely located edge units~\cite{pahl2015containers}. 
These groups, \emph{i.e.}, the \emph{fog}, are closer to the edge than classic cloud providers, allowing for shorter network round-trips and potentially exploitable for latency-sensitive systems.
The fog is more resilient to (local) faults, allows for low latency, location awareness and heterogeneity. 
Devices such as Raspberry Pi~\cite{rapi3b} are popular in this context, due to the low cost and relative high performance to run full-fledged operating systems.
Its energy footprint is orders of magnitude smaller than server-grade machines~\cite{ahvar2019estimating}.
Finally, CPU vendors recently introduced trusted execution environments (TEEs) in small commodity devices as Intel SGX in compute sticks~\cite{cs_sgx} or ARM TrustZone-enabled processors in low-cost devices~\cite{arm_trustzone}.
Hence, fog deployments are considered a viable alternative even for systems with stringent security requirements.

\begin{figure}[!t]
	\centering
	\includegraphics[scale=0.50]{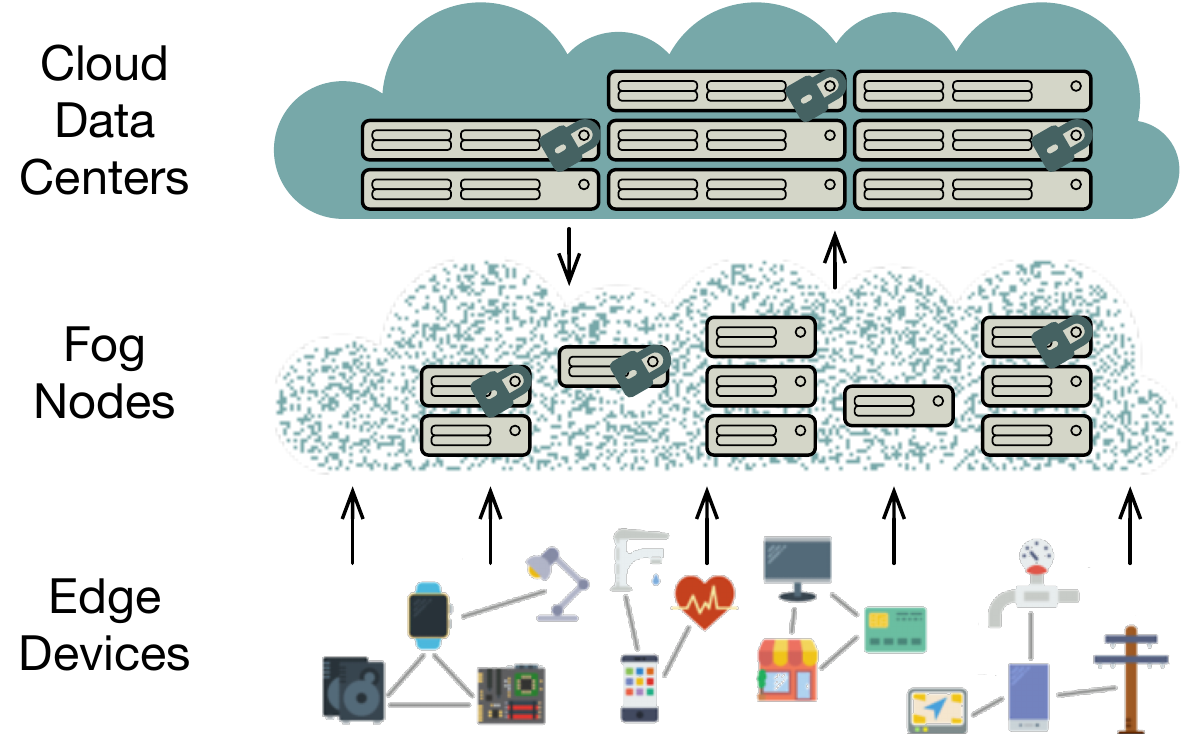}
	\caption{The \sys hierarchical deployment.}
	\label{fig:hierarchy}
	\vspace{-10pt}
\end{figure}

The orchestration of applications deployed across the edge up to the cloud opens several challenges.
Management tools exist to cope with heterogeneous clusters \cite{k8s, mesos}, and yet the specific problems that typically occur in such environments are largely underestimated.
Communication privacy is all the more important as fog devices are primarily connected through wireless networks.
Security in general is important when coping with secure device authentication in fog computing at different gateways, or when dealing with privacy-sensitive data.
Finally, network management and potential performance bottlenecks (which increase latency) currently hinder a larger adoption of fog computing.

There exist solutions~\cite{bellavista2017feasibility, wobker2018fogernetes} that extend well-known infrastructure management platforms (as Docker Swarm, Kubernetes or Apache Mesos) using deployment hooks~\cite{k8sedge}. 
These solutions expose the devices at the edge (and fog) as infrastructure nodes.
However, they do this disregarding some of the critical matters mentioned earlier.


\sys\footnote{\sys is the Roman Goddess of outward journeys.} is a 3-layer hierarchical deployment and monitoring architecture, from the edge (bottom), through the fog (middle) and to the cloud (up). 
Figure~\ref{fig:hierarchy} shows its architecture. 
The smallest manageable resources are the ones deployed at the edge. 
Devices such as Internet gateways in the user's home or sensor data aggregators are typically framed in the edge category. 
We assume these devices to have reliable channels towards the upper stage, whilst they may be connected to sensors through less reliable ones.
At the fog level we consider small sets of rather small devices that can aggregate and process data originating from the edge.
Finally, cloud-based nodes are hosted on public infrastructure, potentially untrusted, with the capacity to easily scale according to the demand.
In this context, \sys continuously monitors several execution metrics for each of the tasks being executed at the different levels of the architecture.
It can then decide to migrate tasks from the edge to the fog and beyond, according to the target optimization functions.
Notable example optimizations supported by \sys include shortest possible execution time, highest security guarantees and smallest energy footprints.

This paper is organized as follows.
The system model is presented in \S\ref{sec:background}, followed by \sys use-cases (\S\ref{sec:usecases}).
We present \sys architecture in \S\ref{sec:architecture}.
We show by means of microbenchmarks (\S\ref{sec:evaluation}) how to achieve higher energy efficiency with a larger number of fog nodes.
A short summary of related work is given in \S\ref{sec:rw}.
Finally, we conclude and sketch our future work in \S\ref{sec:conclusion}. 
\vspace{-5pt}
\section{Model}
\vspace{-5pt}
\label{sec:background}


The \sys system model comprises a 3-layer hierarchical infrastructure. These three layers are introduced as follows.

\textbf{Edge.} Nodes with devices that coordinate to perform a joint or common task. Examples include general processing, computing, or storage tasks, according to the combined capacity of the underlying computing devices.

\textbf{Fog.} Middle layer formed by a set of nodes consisting of low-energy devices. At the same time, they feature sufficient computing capabilities to host tasks that the edge nodes would only complete beyond practical terms. Additionally, fog nodes can be exploited for tasks requiring fast response time (\emph{e.g.}, below 1 sec).

\textbf{Cloud.} Highest level in the architecture, which also provides the highest computing capabilities.
It usually consists of server-grade machines. When the lower layers cannot perform some tasks, these are off-loaded to the cloud nodes.

The layered model as proposed allows the applications to benefit from the combined advantages of each level.
Besides, it also increases the range of opportunities for migrations which in turn allows to better
optimize the overall energy consumption while still respecting the tasks' requirements in terms of  resources and deadlines.

\vspace{-5pt}
\section{Use-cases}
\label{sec:usecases}
\vspace{-5pt}
The model presented in \S\ref{sec:background} allows for a wide range of applications to benefit from the \sys approach.
Below, we shortly describe two of them which we plan on using for further evaluation of our system.
We stress how both use-cases are real application scenarios in the context of the H2020 project LEGaTo.\footnote{\url{www.legato-project.eu}}

\textbf{Use case I - Energy Management.}  An application for collecting the energy consumption of a given place, analyzing it and providing the user with different insights on their consumption and costs.
For instance, the user could have access to a dashboard where all the aggregated data is displayed, the option of setting alarms to be triggered when metrics go above a certain threshold ({e.g.}, the instantaneous overall consumption is higher than a given desired value), and support for nonintrusive appliance load monitoring (NIALM).

\textbf{Use case II - Smart Mirror.} An application for displaying data on a looking-glass.
The application consists of a center display where many software gadgets could be added.
Among those are face and emotion recognition, gesture and speech recognition, object detection, and geolocation for displaying weather or other types of local information.

\begin{figure}[!t]
	\centering
	\includegraphics[scale=0.48]{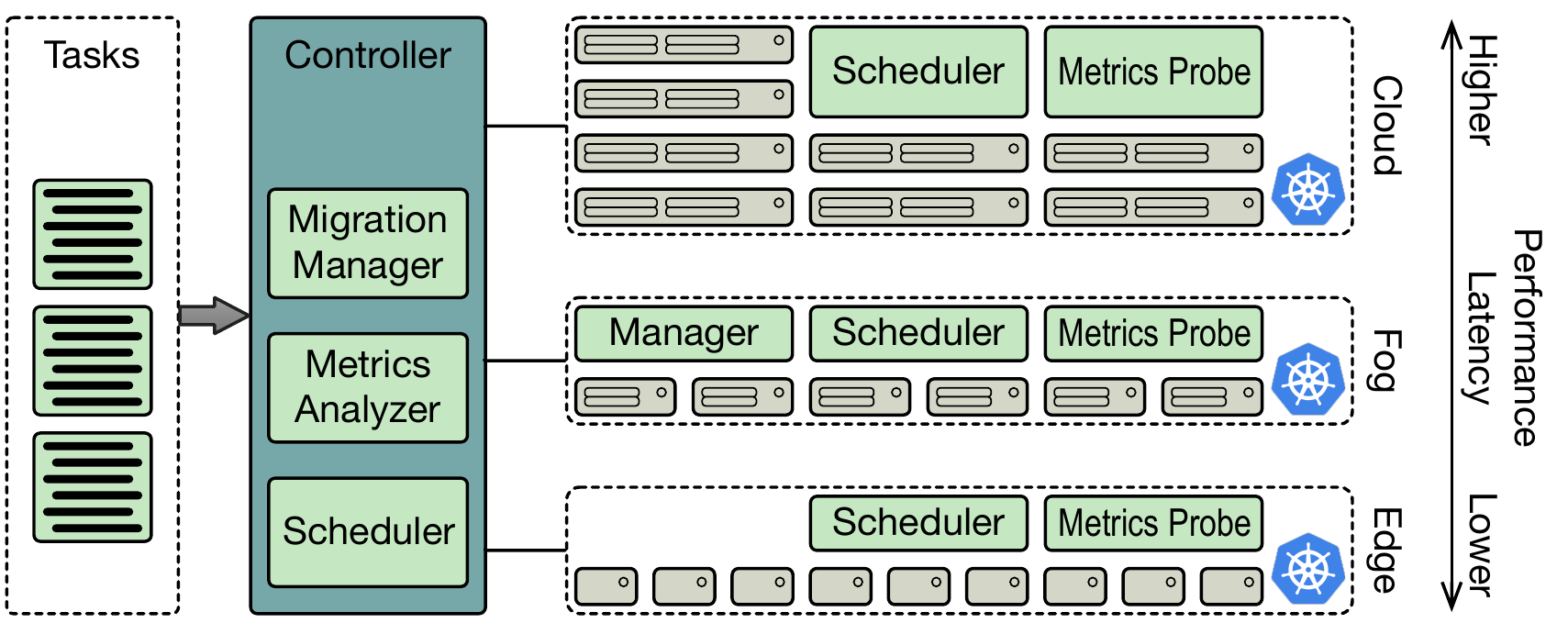}
	\caption{\sys architecture scheme.}
	\label{fig:arc_scheme}
	\vspace{-15pt}
\end{figure}

\vspace{-8pt}
\section{Architecture}
\label{sec:architecture}
\vspace{-8pt}
Figure~\ref{fig:arc_scheme} depicts the architecture scheme of \sys.
The~system closely matches the model described in \S\ref{sec:background}, with the mentioned 3 main layers, namely edge, fog and cloud.
Each layer is modelled as an individual cluster, while at the same time being globally federated and managed by a single controller.
Therefore, each layer can host tasks following a layer-bounded scheduling strategy.
It means that scheduling strategies applied locally might differ from each other.
Each layer hosts a metrics probe, constantly monitoring each node and collecting resource usage metrics (\emph{e.g.}, power, cpu, memory and network usage) as well as life cycle events.

The fog layer embeds an extra, custom manager to take care of analyzing the current configuration of the cluster and the collected metrics, to act upon triggering events.
This manager is responsible for optimizing the overall utilization of the cluster while still responding to individual tasks requirements.

Finally, the proposed \sys architecture uses a controller component which pilots a metrics analyzer, a migration manager and a (global) scheduler.
As the controller holds an overview of the whole system, it can efficiently place new tasks as well as perform migrations into and across the layers.

\vspace{-14pt}
\section{Preliminary Evaluation}
\vspace{-6pt}
\label{sec:evaluation}

We perform microbenchmarks with two different applications running at the fog level, with different numbers of nodes.
Our goal is to show that applications with different requirements might have different optimal deployment settings, and that  our controller can leverage this knowledge to perform the desired optimizations.

\textbf{Evaluation settings.}  The fog layer consists of a Kubernetes cluster with 3~worker nodes.
Each node is a Raspberry Pi 3B+ with the following configuration: ARM Cortex-A53 with \textsc{big.LITTLE} architecture, 4 cores clocked at 1.4GHz, 5W TDP and 1GiB of RAM. 
Each Raspberry Pi is plugged to a PowerSpy device~\cite{powerspy},
which is constantly being monitored and having the measurements stored in a time series database (\emph{i.e.}, InfluxDB) that we mine for power measuring analytics.

The energy consumption of a task $t$ in a cluster of $n$ nodes is given by
$$E(t) =  \sum_{i = 1}^{n} E_{n_i}(t)$$ where $E_{n_i}(t)$ is the 
trapezoidal integral of the power consumption in node $n_i$ during the runtime of task $t$.
We consider two microbenchmarks, detailed next.

\textbf{Application I: Advanced Encryption Standard (AES).} 
\textit{AES}~\cite{daemen2013design} is a symmetric-key cryptographic key block cipher scheme based on the substitution–permutation network principle.
We leverage \textit{PyAES}~\cite{pyaes}, an open source Python implementation of AES.
The benchmark performed encrypted 92000 bytes of text with 128 bit key for 243~iterations.

\textbf{Application II: PageRank.}
The PageRank algorithm is a well-established approach that evaluates a weighted quantity of links to a webpage to determine a its relative score of importance.
We implemented (and released as open-source) \textit{PyPR}~\cite{pagerank}, our own python implementation of the PageRank algorithm.
For this benchmark, we execute 10 iterations per page over Google's web graph~\cite{leskovec2009community}.

\begin{figure}[t!]
	\centering
	\includegraphics[scale=0.95]{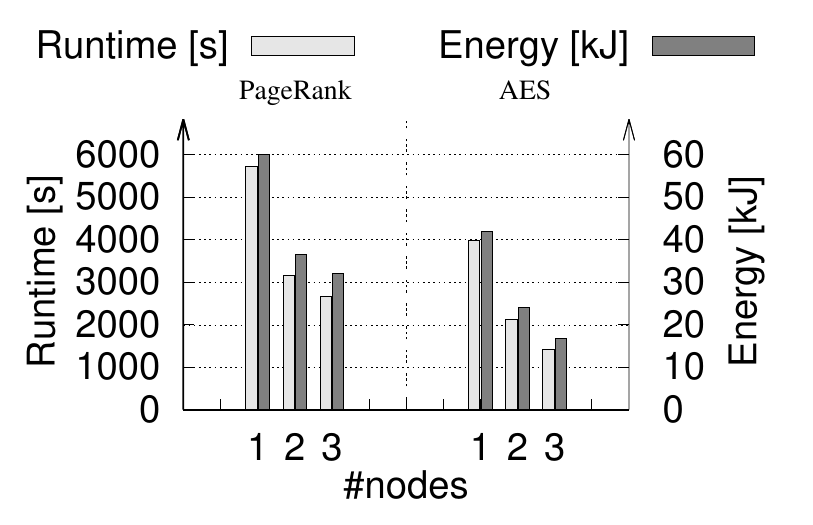}
	\vspace{-10pt}
	\caption{Runtime \emph{vs.} energy consumption of applications \textit{Advanced Encryption Standard  (AES)} and \textit{PageRank} using different number of nodes in the fog.}
	\label{fig:preliminary_evaluation}
	\vspace{-18pt}
\end{figure}

\textbf{Results.}
The two application benchmarks are executed both sequentially and in parallel.
For the latter, we distribute the computation across two or three nodes.
Figure~\ref{fig:preliminary_evaluation} shows the runtime and energy results of both applications.
The runtime represents the execution time of the sequential execution and the makespan of both parallel executions.
The energy is calculated as the overall sum of the energy spent, with one, two or three nodes.

Analyzing the results of both applications, we observe a considerable decrease in the runtime and energy consumption as the number of nodes increase.
These preliminary results suggest that, for some applications, more nodes lead to better efficiency, both in terms of performance and energy.

\vspace{-5pt}
\section{Related Work}
\vspace{-5pt}
\label{sec:rw}

Many applications could benefit from the infrastructure model we described.
A distributed recommendation deployed from the edge to the cloud in Cagliari (Italy) airport~\cite{salis2018anatomy} follows a similar model.
Their application offers the users with recommendations based on their geolocation in the airport and the user's interests, tuned by user' feedback.
The deployment is distributed, where the users applications are in the edge, an aggregator component in the fog, and finally, a component performing machine learning tasks in the cloud.

A similar approach is presented in~\cite{masip2016foggy}: their solution suggests layer management must be coordinated.
\sys considers the management of a distributed layered computing system, as well as energy and security matters.

\vspace{-5pt}
\section{Conclusion and Future Work}
\vspace{-5pt}
\label{sec:conclusion}

We presented \sys, a monitoring and scheduling architecture for edge-to-cloud computing.
This system monitors key performance metrics, such as energy consumption at the edge, and can trigger the migration of tasks towards fog or cloud computing nodes.
Our early results demonstrate that task migrations can effectively lead to energy savings even when the number of nodes increases.

We intend to further investigate the effects observed in our evaluation including a more diverse set of nodes and workloads.
We will investigate the impacts of {CPU} frequency scaling in our results, and how to exploit that in \sys.

We plan to support task migration across heterogeneous nodes within and across layers whenever a better suited deployment setting is detected or predicted.

Finally, we will investigate security aspects within \sys, for applications operating on private data, by leveraging a mix of Intel SGX and ARM~TrustZone nodes.

\vspace{-5pt}
\section*{Acknowledgements}
\vspace{-5pt}
\label{sec:ack}
The research leading to these results has received funding from the European Union's Horizon 2020 Programme under the LEGaTO Project (www.legato-project.eu), grant agreement No$^{\circ}$ 780681.
\vspace{-8pt}

\printbibliography

\end{document}